%% file: main.tex
\documentclass[sigconf, 10pt, nonacm]{acmart}
\usepackage{epsfig,xspace,url,xcolor}
\usepackage{subcaption}
\usepackage{amsmath}
\usepackage{mathtools}
\usepackage{amsthm}
\usepackage{tikz}
\usepackage[most]{tcolorbox}
\usepackage{caption}

\usepackage[small,compact]{titlesec}
\titlespacing*{\section}{0pt}{4pt}{4pt}%
\titlespacing*{\subsection}{0pt}{4pt}{4pt}%
\titlespacing*{\subsubsection}{0pt}{0pt}{0pt}%

\usepackage{graphicx}
\usepackage[ruled, lined, linesnumbered, commentsnumbered, longend]{algorithm2e}
\usepackage[noend]{algpseudocode}
\usepackage{url}
\usepackage{colortbl}
\usepackage{thmtools}
\usepackage{thm-restate}

\usepackage{wrapfig}
\usepackage[subtle]{savetrees}
\usepackage[framemethod=TikZ]{mdframed}
\usepackage{siunitx}

\renewcommand\footnotetextcopyrightpermission[1]{}

\input{macros}

\let\ACMmaketitle=\maketitle
\renewcommand{\maketitle}{\begingroup\let\footnote=\thanks \ACMmaketitle\endgroup}
\let\height\relax

\begin{document}

\sloppy
\newcommand{\name}{\textsc{name}\xspace}

\title{What Is Reconfiguration Delay, Really?\\An End-to-End View of Photonic Switching}

\author{Deeksha P Rao}
\affiliation{
  \institution{Purdue University}
}
\author{Vamsi Addanki}
\affiliation{
  \institution{Purdue University}
}
\renewcommand{\shortauthors}{Deeksha P Rao and Vamsi Addanki}

\begin{abstract}
Reconfiguration delay is a defining systems parameter of photonic switching. Yet, it is almost universally modeled as a single constant in network design and circuit scheduling algorithms, namely the optical switching time reported by the device. We argue that this abstraction is no longer sufficient. From an application's perspective, reconfiguration begins when a new circuit is requested and ends only when communication resumes over the new path. The latency observed by the application spans an end-to-end critical path across multiple hardware and software components, not just the optical switch.

This paper asks a simple question: what is reconfiguration delay, really? We argue that it should be understood as an end-to-end systems property rather than a device property. To make this concrete, we dissect the critical path of a circuit transition, spanning host-to-controller communication, controller processing, electrical signaling, and optical switching. Our measurements show that optical switching itself accounts for only a small fraction of the end-to-end latency, even before considering transceiver recovery and receiver-side reactions. This perspective raises a fundamental systems question: if most of the end-to-end delay lies outside the switch fabric, what should future photonic switching systems optimize? We conclude by outlining a research agenda for reducing end-to-end reconfiguration delay across the hardware--software stack.
\end{abstract}

\maketitle
\pagestyle{empty}

\section{Introduction}

The growing demand for network bandwidth, together with the slowdown of Moore's Law, has renewed interest in photonic switching, both in academia~\cite{Harvest,harness,kumar2025pccl,10.1145/3772356.3772395,10.1145/3712285.3759842,Scheduling_parallel_OCS,addanki2025vermilionl,10.1145/3718958.3750465,10.1145/3779212.3790238,10.1145/3579371.3589350,torrijos2026industry} and industry~\cite{iPronics2025,9007742}. Photonic switching promises high bandwidth and low power consumption, making it an attractive communication substrate for datacenters and high-performance computing~\cite{10.1145/3748273.3749203,10.1145/3779212.3790238,10.1145/3712285.3759842,Harvest,10.1145/3579371.3589350, Opus}, particularly for chip-to-chip communication in scale-up systems~\cite{10.1145/3772356.3772395,10.1145/3696348.3696856}.

The circuit-switched nature of photonic interconnects introduces two fundamental systems questions: \first what circuit topology should be established, and \second when should the topology be reconfigured? Both questions fundamentally depend on the reconfiguration delay of the interconnect~\cite{Harvest,10.1145/2896377.2901479,10.1145/3772356.3772395,10.1145/2716281.2836126,10.1145/3712285.3759842}. Existing works typically model reconfiguration delay as a single constant determined by the switching technology, while device vendors and photonic hardware research report the intrinsic optical switching time, ranging from tens of nanoseconds for electro-optic switches~\cite{Lu:16,qiao201732} to milliseconds for mechanical switches~\cite{polatis}. This paper asks a different question.

\begin{figure}[t]
    \centering
    \includegraphics[width=0.75\linewidth]{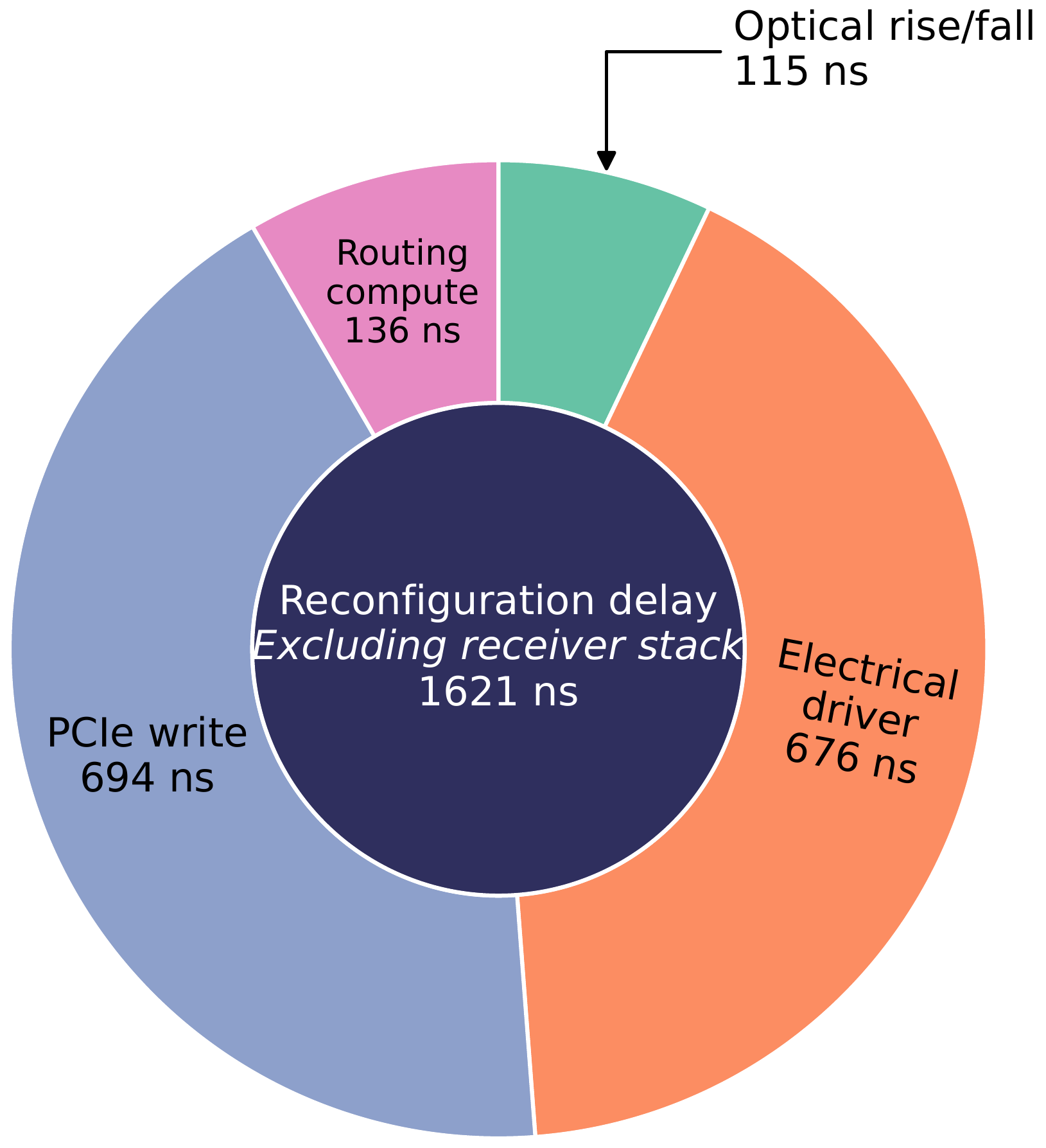}
    \caption{Reconfiguration delay breakdown excluding receiver locking, with optical rise/fall time contributing only $7\%$.}
    \label{fig:reconf_delay}
    \vspace{-0.7cm}
\end{figure}

\begin{quote}
\emph{What does reconfiguration delay actually comprise in a photonic switching system?}
\end{quote}

To answer this question, we consider synchronous, source-initiated photonic switching, in which a traffic source (e.g., a GPU) requests a new circuit and a controller programs the optical interconnect~\cite{10.1145/3718958.3750465,Harvest,10.1145/3748273.3749203}. We distinguish two intervals: the control path delay, from circuit request to light arrival, and receiver recovery, from light arrival to traffic resumption. Figure~\ref{fig:reconf_delay} traces the resulting end-to-end critical path in our testbed using nanosecond-scale electro-optic switches. A striking observation is that optical switching accounts for only $\approx7\%$ of the measured reconfiguration delay, even before considering the well-known problem of transceiver locking at the receiver. It covers the control-path delay ($\approx$[1.6]\,$\mu$s in our prototype); receiver recovery is about six orders of magnitude larger on commodity NICs ($\approx$1\,s) and is analyzed separately in \S\ref{subsec:receiver_latency}.

The control-path latency arises from stages that are rarely considered together. Before the optical switch can change state, the source must communicate the circuit request to the controller. Using a GPU connected to an FPGA controller over PCIe, this communication alone contributes $\approx694$\,ns on average, making it the largest component in our testbed. The controller must then translate the requested circuit into switch configurations. In our FPGA implementation, routing an $8\times8$ Bene\v{s} interconnect requires $\approx136$\,ns, increasing to $1.3\,\mu$s for a $32\times32$ interconnect, nearly an order of magnitude larger than the optical switching time itself. Finally, the control signal must propagate from the controller to the optical switch before the optical transition occurs, introducing another non-negligible delay. Once the circuit has been established, the destination transceiver must reacquire lock, a well-known challenge that can add tens of microseconds~\cite{forencich2020system}. We discuss this receiver-side latency separately and exclude it from the end-to-end measurements presented in Figure~\ref{fig:reconf_delay}. Importantly, each individual latency, including the optical switching time itself, are neither constant or deterministic, and introduce non-negligible variance when modeling nanosoecond-scale systems. We elaborate on each of these components in Section~\ref{sec:dissecting} and quantify their contribution to the overall reconfiguration delay.

This perspective shifts the optimization target from the intrinsic switching time of the optical device to the end-to-end latency experienced by distributed applications. The purpose of this paper is not to claim that these individual delays are new. Rather, we place them on a common timeline to expose the critical path of photonic switch reconfiguration. Naturally, from a network design and circuit scheduling perspective, we urge the community to consider variance in reconfiguration delay as a first-class parameter, opening new avenues for research in online algorithms where the cost of reconfiguration is partially known and only revealed after the fact. From a systems and hardware perspective, we believe it is high-time to revisit the classic problem of interconnect routing e.g., Bene\v{s} routing, and efficient algorithms that can compute switch states within a small fraction of the switching time. We discuss several future research directions at the end of the paper.

\begin{figure}[t]
\centering
\includegraphics[width=0.95\linewidth]{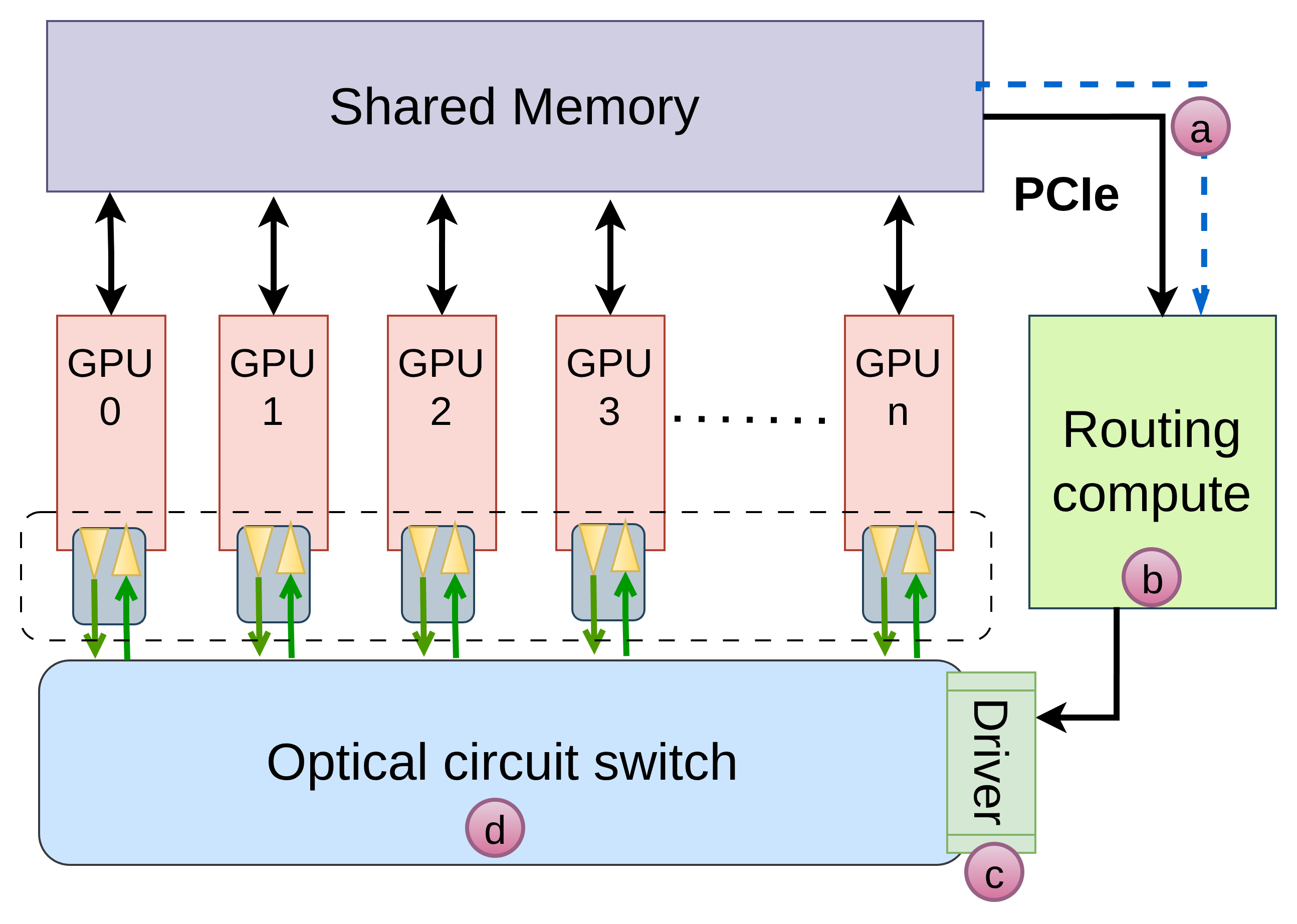}
\caption{Hardware prototype}
\vspace{-4mm}
\label{fig:prototype}
\end{figure}

\section{Hardware Prototype}

As part of our broader effort to bring photonic interconnects closer to GPUs and end hosts in modern scale-up systems, we developed an end-to-end hardware prototype to study reconfiguration delay.

Figure~\ref{fig:prototype} illustrates the architecture. Our prototype consists of eight NVIDIA RTX PRO 6000 GPUs, each connected to a BlueField-3 DPU over PCIe Gen5 with dedicated PCIe bandwidth. Since current GPUs cannot directly interface with optical transceivers, we treat the GPU, DPU, and optical transceiver as a single compute node. The BlueField-3 hosts the optical transceiver and operates in DPU mode, allowing forwarding rules to be programmed directly into hardware through Open vSwitch. This abstraction affects only the data path; the control path, which is the focus of this paper, remains unchanged. Each transceiver connects to the photonic interconnect through single-mode LC fiber.

The photonic interconnect is an $8\times8$ Bene\v{s} network constructed from seventeen $2\times2$ electro-optic switches~\cite{10.1145/321439.321449}. Each switch consists of a birefringence crystal whose refractive index changes in response to an applied voltage~\cite{borshch2013nanosecond,shen2014electro}, thereby changing the optical path. Consequently, every optical reconfiguration requires an electrical driver circuit that controls the voltage to transition each switch between its cross and bar states.

An FPGA acts as the reconfiguration controller between the compute nodes and the optical fabric. Connected to the GPUs over PCIe Gen5 and to the optical switches through GPIO, it receives circuit requests, computes the corresponding Bene\v{s} switch configuration, and simultaneously drives all seventeen switches using TTL control signals. This control path naturally decomposes reconfiguration into four stages: communicating the circuit request from the GPU to the FPGA, computing the switch configuration, propagating control signals to the switch drivers, and performing the optical transition.

Our prototype supports arbitrary communication topologies across the eight GPUs while allowing circuits to be reconfigured at runtime. More importantly, it exposes every stage of the end-to-end reconfiguration pipeline, allowing us to isolate and quantify the latency introduced by each stage rather than observing only the optical switching event.

\section{Dissecting the Reconfiguration Delay}
\label{sec:dissecting}

Figure~\ref{fig:prototype} illustrates the control path responsible for establishing a new optical circuit. Every reconfiguration begins when the communication runtime determines that a new topology is required, for example, at the boundary of a collective communication step. The requested logical topology is communicated from the compute node to the switch controller, translated into individual switch configurations, and finally propagated as electrical control signals that actuate the optical switches. While the optical transition itself lasts only a few nanoseconds, every stage preceding it contributes to the end-to-end reconfiguration delay.

This section examines the control path corresponding to points \circled{a}--\circled{d} in Figure~\ref{fig:prototype}. Our controller is implemented on an AMD VCK190 Versal FPGA connected to the compute nodes through PCIe Gen5 and to the optical switches through GPIO. Upon receiving a circuit request, the FPGA computes the corresponding Bene\v{s} switch configuration and simultaneously drives the electrical interfaces of all optical switches. We next quantify the latency introduced by each stage of this pipeline.

\subsection{Host-to-Controller Communication}\label{pcie}

The first stage of reconfiguration is communicating the circuit request from the compute node to the FPGA controller. The VCK190 integrates the Cache Coherent PCIe Module 4 (CPM4), which supports several PCIe communication mechanisms including Programmable I/O (PIO), Bus Master DMA (BMD), XDMA, QDMA, and AXI bridges. Since a circuit request consists of only a few bytes describing the desired topology, throughput is secondary to latency. We therefore employ PIO, which exposes the FPGA BAR space directly to the host and allows the GPU to access FPGA registers with minimal software and hardware overhead. In contrast, DMA-based approaches incur additional setup costs that dominate such small control-plane transactions.

\begin{figure}[!h]
    \centering
    \includegraphics[width=0.75\linewidth]{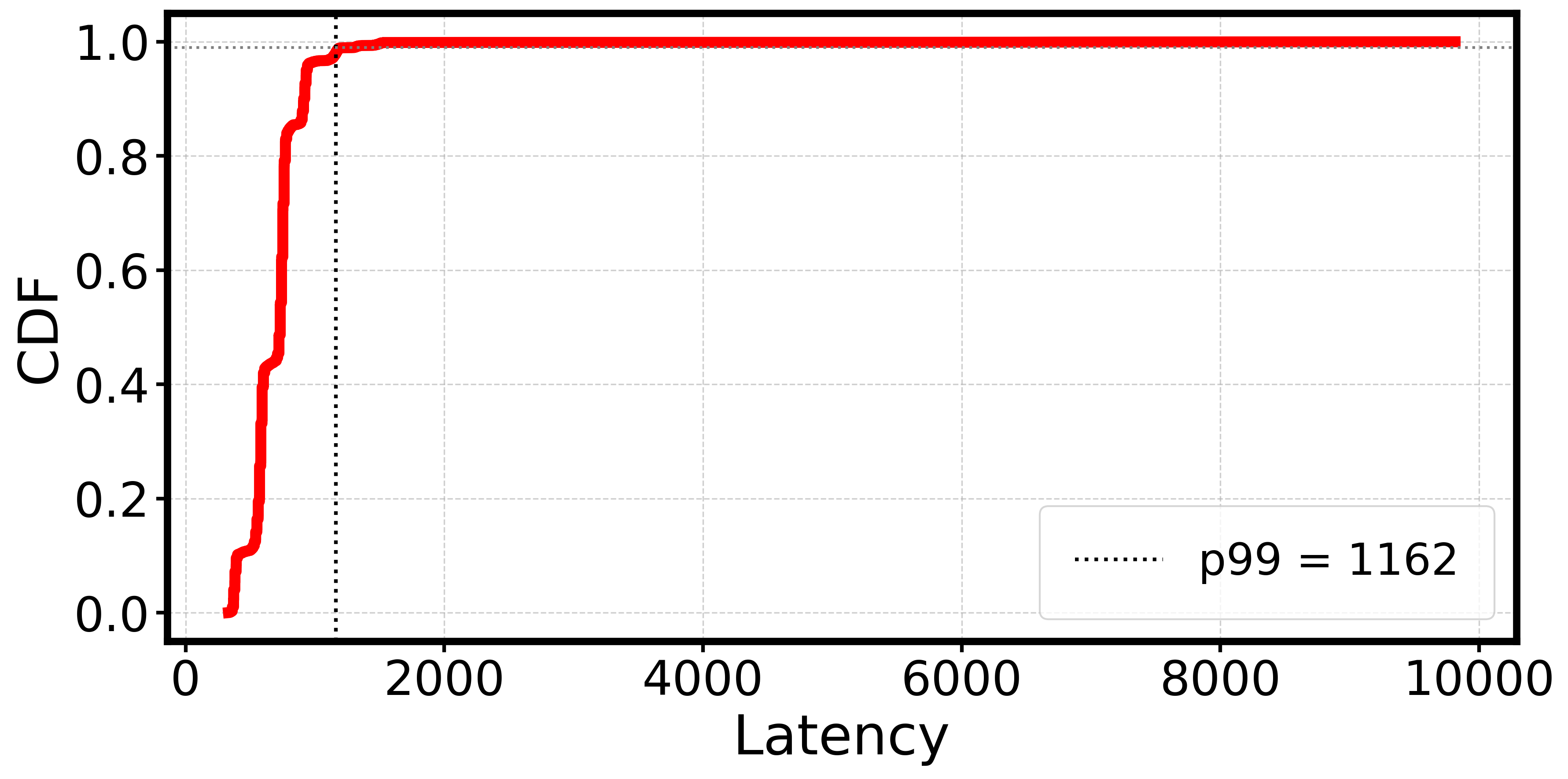}
    \caption{CDF of PCIe write latency between the compute node and the FPGA controller, with $99^{th}$ percentile latency of  $1162$\,ns and average latency of $694$\,ns.}
    \label{pcie_delay}
    \vspace{-2mm}
\end{figure}

Figure~\ref{pcie_delay} shows the cumulative distribution of PCIe write latency measured between the host and the FPGA. The distribution rises sharply between $330$\,ns and $1\,\mu$s, with an average write latency of approximately $694$\,ns. More than $99\%$ of writes complete within $1.15\,\mu$s, making PCIe communication the single largest contributor to the end-to-end reconfiguration delay in our prototype.

\subsection{Controller Processing}\label{routing}

Once the circuit request reaches the FPGA, it must be translated into individual switch configurations. By routing, we refer to the process of mapping a requested logical topology to the state of every $2\times2$ switch in the underlying interconnect. The complexity of this translation depends on the switching architecture. Optical switching fabric are typically constructed in modular fashion using units of switching elements. Different topologies can be employed to scale an $N\times N$ from a unit $1\times2$ or a $2\times2$. Among all popular topologies include PILOSS~\cite{piloss}, Double-Layer Network(DLN)~\cite{low-crosstalk}, Crossbar~\cite{crossbar} and \benes topologies~\cite{optdatacentre}. Every topology has a different number of input, output, and intermediary layers. With the increase in the number of layers, the dependency between paths reduces. And thus, the partial reconfigurability of the switching fabric increases~\cite{benlee}. However, it directly affects the number of switches used and thus the complexity of mapping  logical topology to the switch states. While crossbar fabrics can establish connections with little or no routing computation, they require $O(N^2)$ switching elements and become prohibitively expensive at scale~\cite{6770468}. Practical photonic interconnects therefore increasingly adopt multistage Rearrangeably Non-Blocking (RNB) fabrics such as Bene\v{s}, which realize arbitrary permutations using only $O(N\log N)$ switching elements at the cost of online routing computation~\cite{opferman,4x4,qiao201732,Lu:16}. Other scalable multistage switching architectures, including Clos and related variants, pose similar configuration problems that must be solved before the optical fabric can be reconfigured~\cite{6770468}. Consequently, controller computation itself becomes part of the critical reconfiguration path.

\begin{figure}[!h]
\begin{subfigure}{0.48\linewidth}
\centering
\includegraphics[width=1\linewidth]{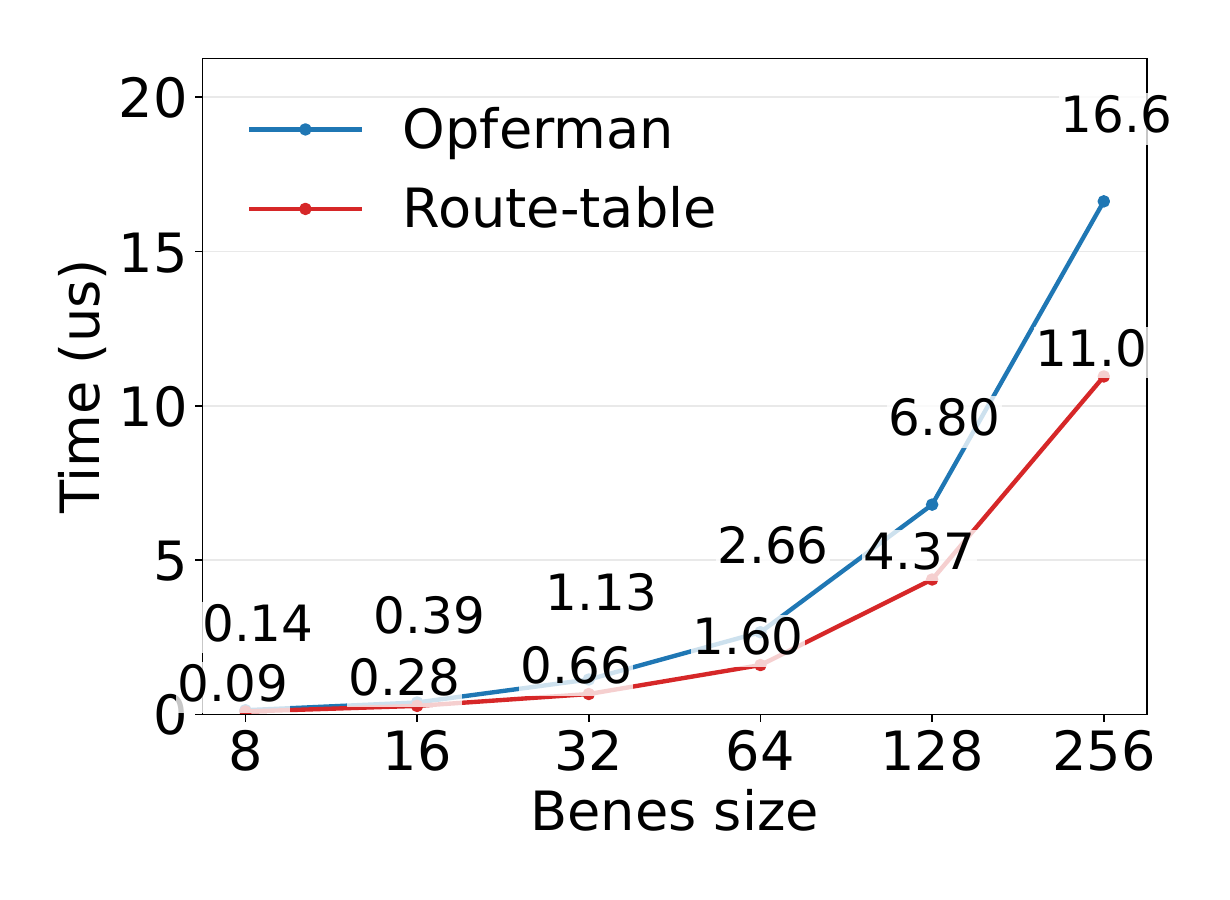}
\caption{Hardware execution time}
\label{fig:benestime}
\end{subfigure}\hfill
\begin{subfigure}{0.48\linewidth}
\centering
\includegraphics[width=1\linewidth]{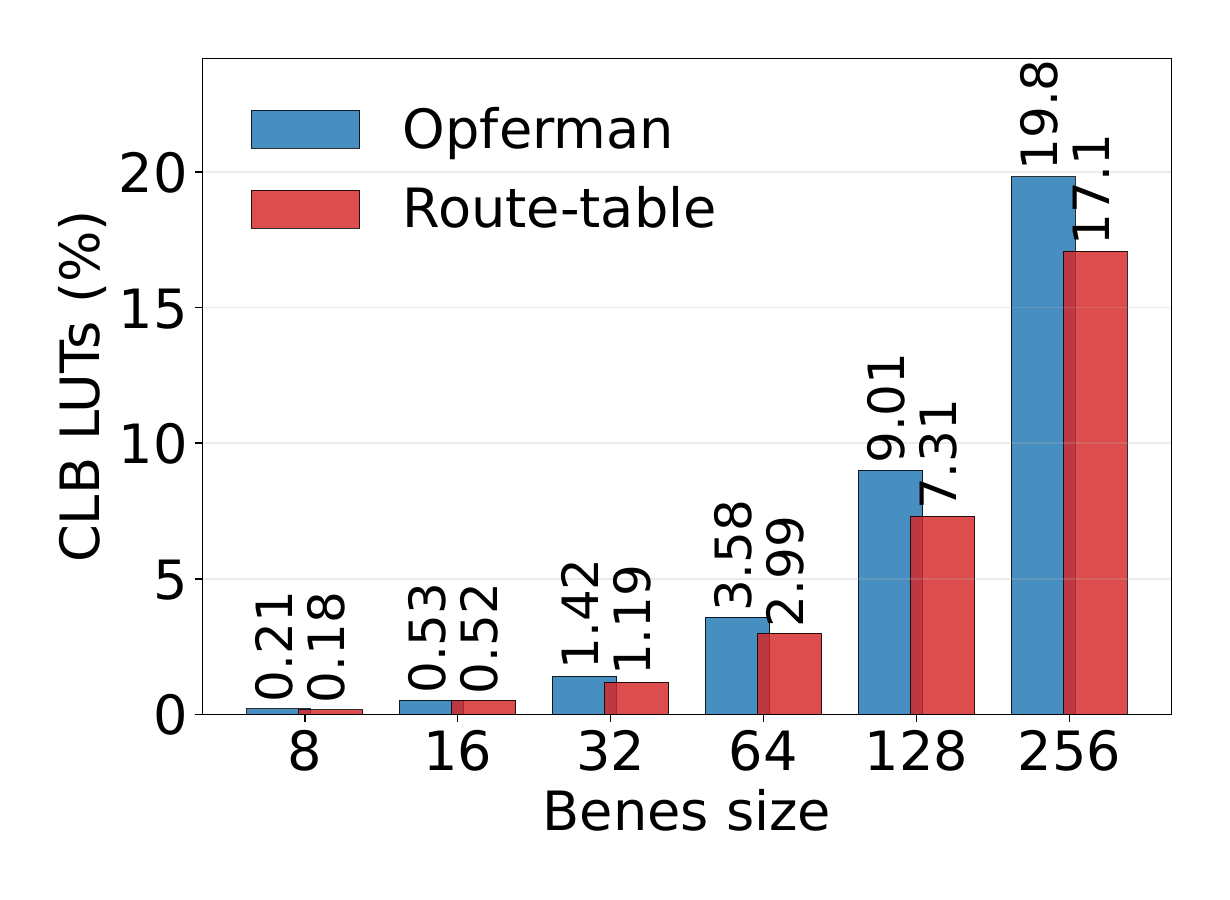}
\caption{FPGA resource utilization}
\label{fig:benes-clb-luts}
\end{subfigure}
\caption{Routing computation time and hardware resource utilization for Bene\v{s} interconnects shows non-negligible overhead in the critical path.}
\label{fig:fpga}
\vspace{-3mm}
\end{figure}

Figure~\ref{fig:benestime} reports the hardware execution time of two FPGA implementations of Bene\v{s} routing: a graph-coloring-based implementation following Opferman's algorithm~\cite{opferman} and a route-table-based implementation~\cite{nikolaidis2022novel}. Figure~\ref{fig:benes-clb-luts} shows the corresponding FPGA resource utilization. For an $8\times8$ interconnect used in our prototype, routing completes in approximately $116$--$160$\,ns, already exceeding the intrinsic optical switching time of many electro-optic devices. As the switch fabric scales, however, routing latency grows rapidly~\cite{benesroute,parallelroute}. For a $256\times256$ Bene\v{s} network, the graph-coloring implementation requires $16.6\,\mu$s, while the route-table implementation requires $11.0\,\mu$s. The graph-coloring implementation also consumes approximately $20\%$ of the FPGA LUTs, compared to $17\%$ for the route-table implementation.

The key observation is not which routing implementation is faster, but that routing itself quickly becomes a first-order contributor to reconfiguration delay. While optical switching technology has steadily reduced device switching times to only a few nanoseconds, controller computation continues to grow with the size and complexity of the switching fabric. As photonic interconnects scale, the latency of computing switch configurations can dominate the intrinsic optical switching time, making controller processing an increasingly important target for end-to-end optimization.

\textbf{When is routing on the critical path?} Many workloads that motivate runtime reconfiguration necessitate online computation: in data-dependent MoE all-to-all traffic~\cite{10.1145/3718958.3750465}, multi-tenant job arrivals, or failures, the next configuration is known only at runtime. This places routing directly on the critical path. However, routing is not always on the critical path: demand-oblivious schedules~\cite{10.1145/3098822.3098838,10.1145/3651890.3672248} and collectives whose communication patterns are known at compile time~\cite{Harvest,kumar2025pccl} allow switch states to be precomputed. We discuss this tradeoff, and the limits of precomputation at scale, in \S\ref{sec:future}.

\subsection{Insertion Loss}

Routing algorithms are often formulated purely in terms of connectivity. In practice, however, the optical properties of the interconnect introduce additional constraints. Among these, insertion loss (IL) is particularly important because it determines whether a candidate path satisfies the available optical power budget.

\begin{figure}[!h]
\centering
\begin{subfigure}{0.48\linewidth}
    \centering
    \includegraphics[width=1\linewidth]{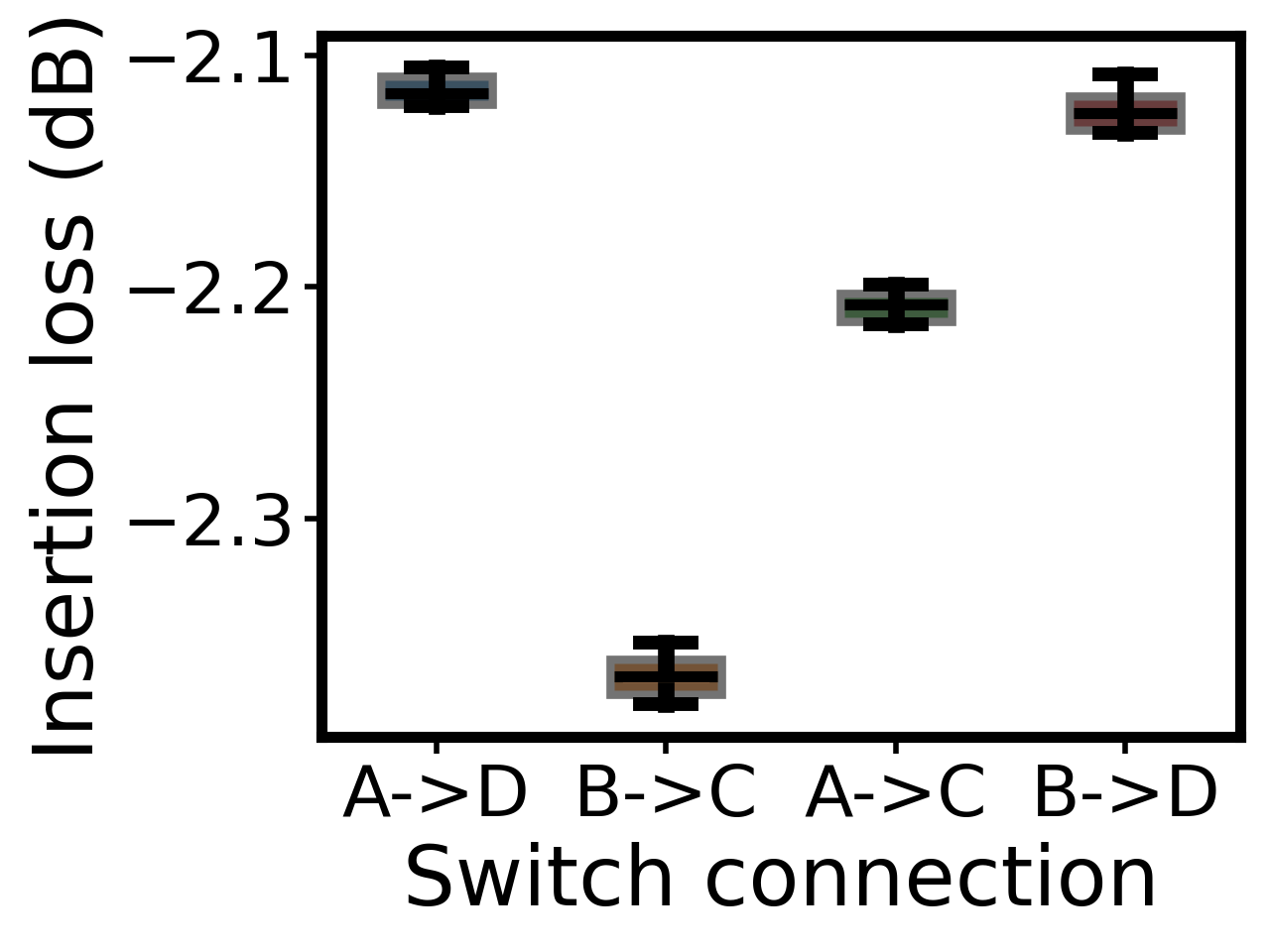}

    \caption{Insertion loss of a switch}
    \label{IL_SW1}
\end{subfigure}\hfill
\begin{subfigure}{0.48\linewidth}
    \centering
    \includegraphics[width=1\linewidth]{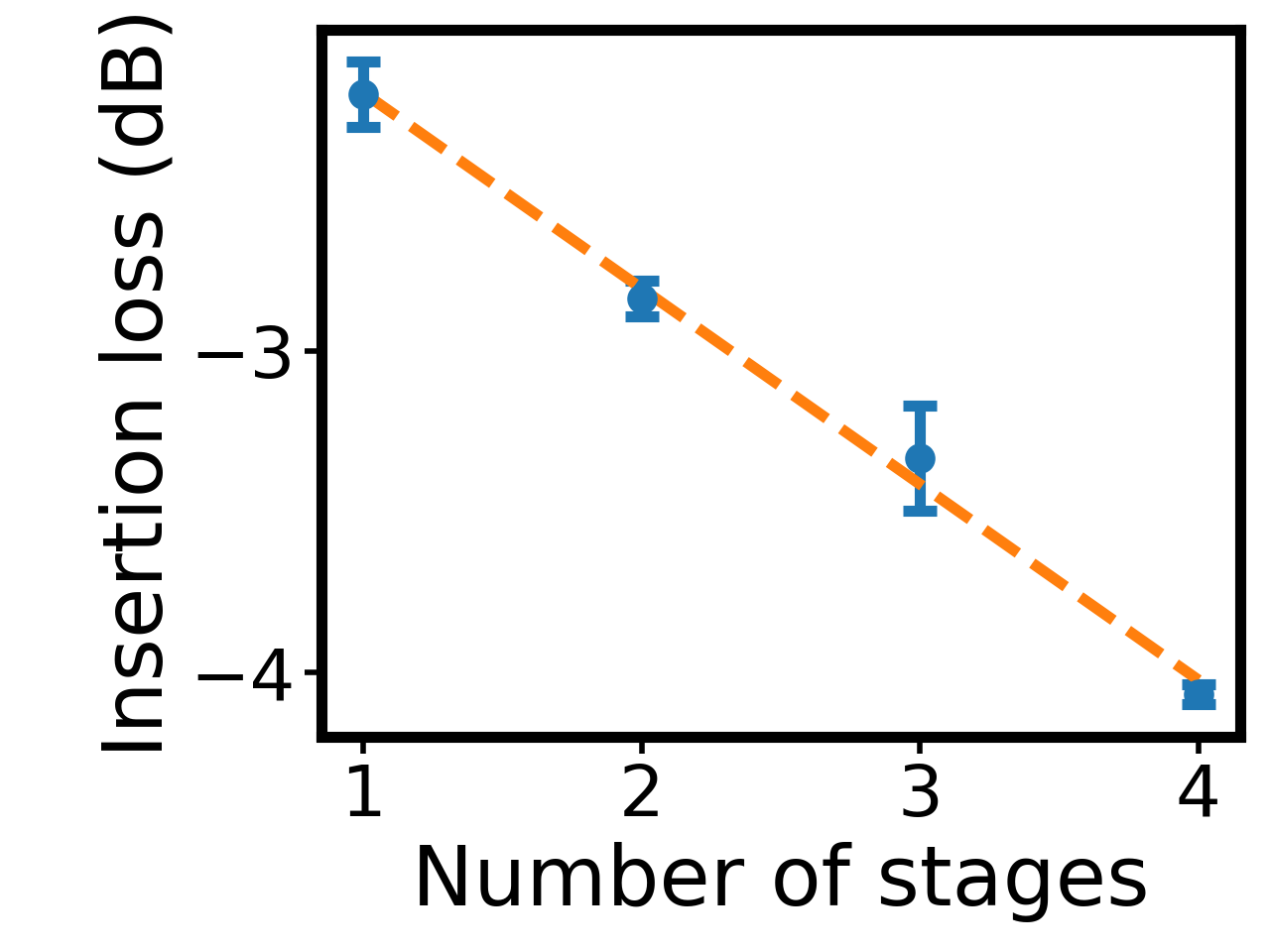}

    \caption{Insertion loss with Scaling}
    \label{IL_range}
\end{subfigure}
\caption{Insertion loss with different switch states and varied number of stages.}
\vspace{-4mm}
\label{fig:IL}
\end{figure}
Insertion loss becomes increasingly important as the switch fabric scales, and its growth depends strongly on the topology connecting the underlying switching elements. The topology determines how many elements each optical path traverses and whether this number varies across source--destination pairs. 
Fortunately, Bene\v{s} networks possess an important property: every input-output path traverses the same number of switching stages. As a result, all source-destination pairs experience nearly identical path loss, allowing routing to be performed without explicitly accounting for insertion loss. Figure~\ref{IL_SW1} shows the measured insertion loss of a representative $2\times2$ switch. Although different switching states exhibit small variations in insertion loss, Figure~\ref{IL_range} shows that the end-to-end insertion loss scales approximately linearly with the number of switching stages. Since every path in a Bene\v{s} network traverses the same number of stages, the resulting path loss remains nearly uniform across all routes.

This property does not generally hold for other photonic switching architectures. Grid-based and irregular topologies admit multiple paths with different lengths and different numbers of switching elements, resulting in different insertion loss for different routes~\cite{gridloss,crossbarloss}. Routing algorithms for such architectures must jointly consider connectivity and optical feasibility , increasing the complexity of controller processing and, consequently, the end-to-end reconfiguration latency.
\subsection{Electrical Signaling}\label{subsec:elecdriver}

The final stage of the control path is actuating the optical switches. This stage bridges the digital control plane and the optical fabric, translating controller outputs into the physical signals required by the switching elements. Every photonic switching technology requires such an interface between the controller and the optical device, making it an unavoidable component of end-to-end reconfiguration delay.

The design of the actuation circuitry depends on the underlying switching technology. MEMS switches require precision actuators and high-resolution DACs to steer micro-mirrors with low settling times~\cite{memsdriver}. Thermo-optic and magneto-optic switches require drivers capable of generating controlled thermal and magnetic fields, respectively. Electro-optic switches instead require high-speed voltage drivers with fast rise and fall times. Although recent electro-optic devices demonstrate intrinsic switching times of only a few nanoseconds~\cite{piloss,3x3,4x4}, and emerging devices project picosecond-scale operation~\cite{picoswitch}, the latency of the driver circuitry often exceeds that of the optical switching element itself. Faster switching devices alone do not necessarily reduce the end-to-end reconfiguration delay experienced by applications.

\begin{figure}[!h]
\begin{subfigure}{0.48\linewidth}
    \centering
    \includegraphics[width=1\linewidth]{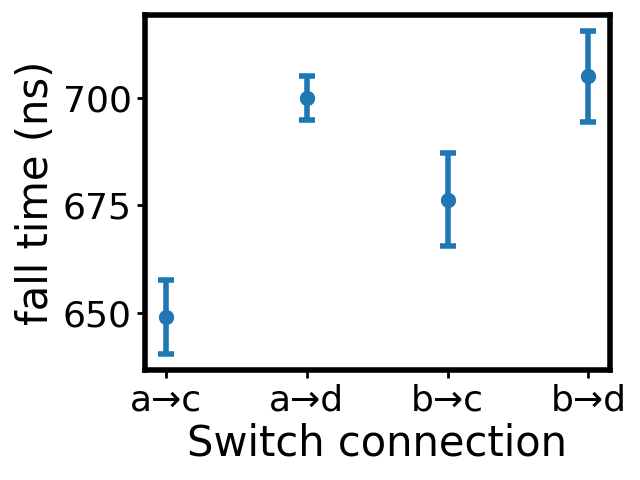}

    \caption{Fall time}
    \label{fig:falltime}
\end{subfigure}\hfill
\begin{subfigure}{0.48\linewidth}
    \centering
    \includegraphics[width=1\linewidth]{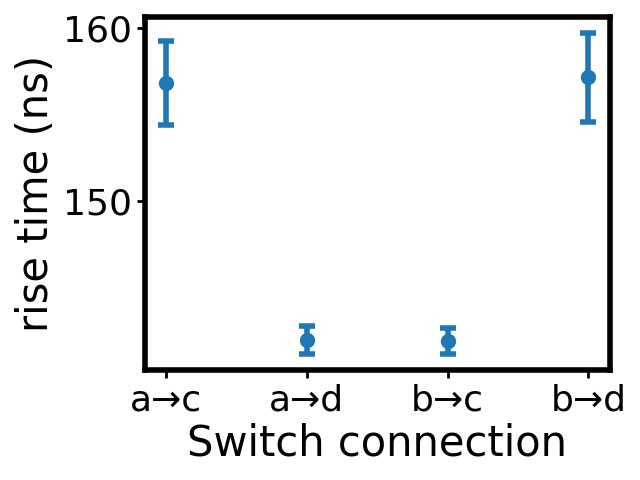}

    \caption{Rise time}
    \label{fig:risetime}
\end{subfigure}
\caption{Electrical-to-Optical rise/fall time for different switch configurations: $a\rightarrow c \ \& \  b\rightarrow d$ represent  bar connection and $a\rightarrow d \ \& \  b\rightarrow c$ represent the cross connection.}
\label{fig:rise-fall-variance}
\vspace{-4mm}
\end{figure}

\begin{figure*}
\centering
\begin{subfigure}{0.45\linewidth}
\centering
\includegraphics[trim=1cm 0 1.5cm 0, clip,width=\linewidth]{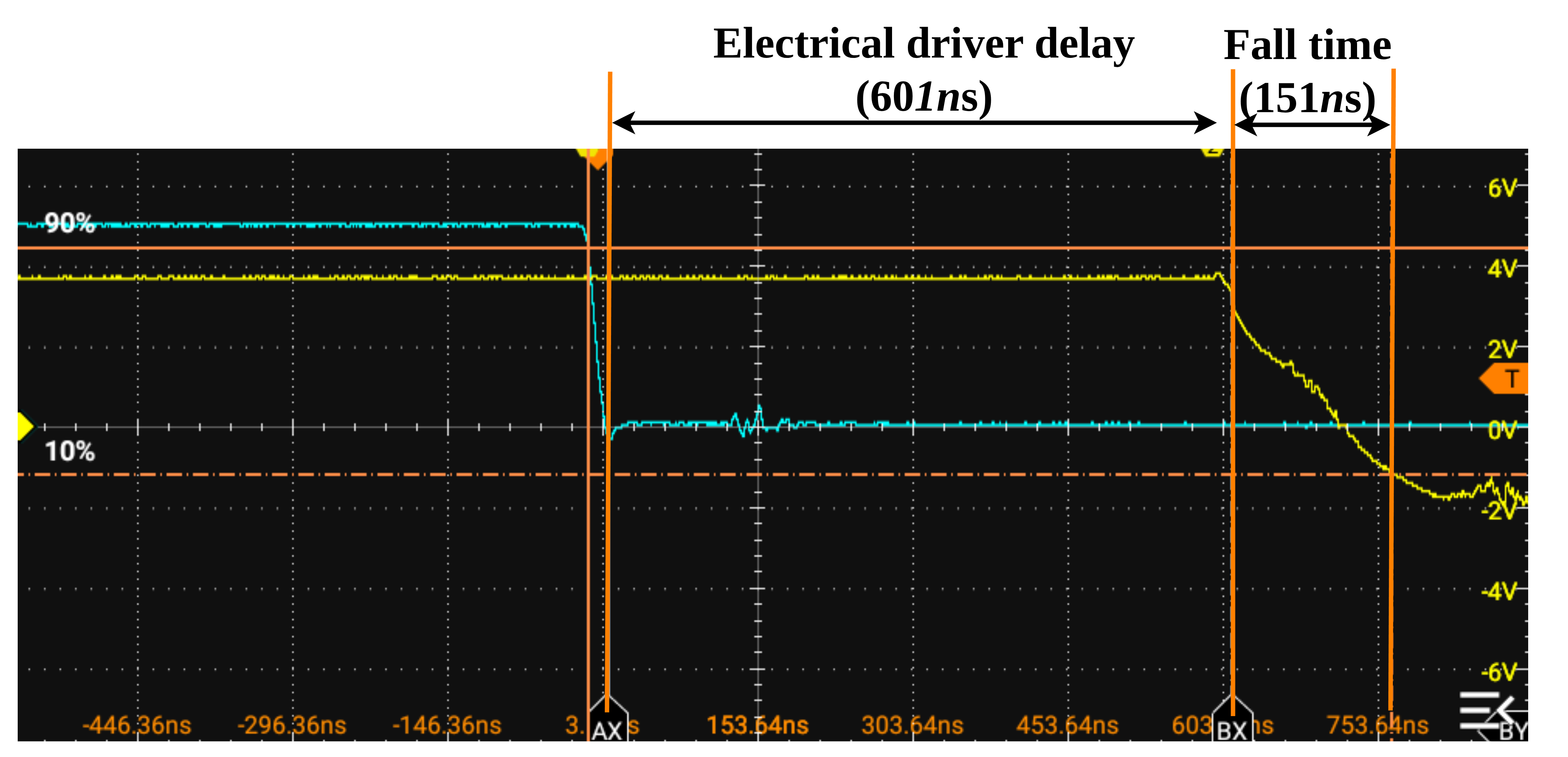}
\caption{Electrical fall to optical fall}
\end{subfigure}\hfill
\begin{subfigure}{0.45\linewidth}
\centering
\includegraphics[trim=0 0 0cm 0, clip,width=\linewidth]{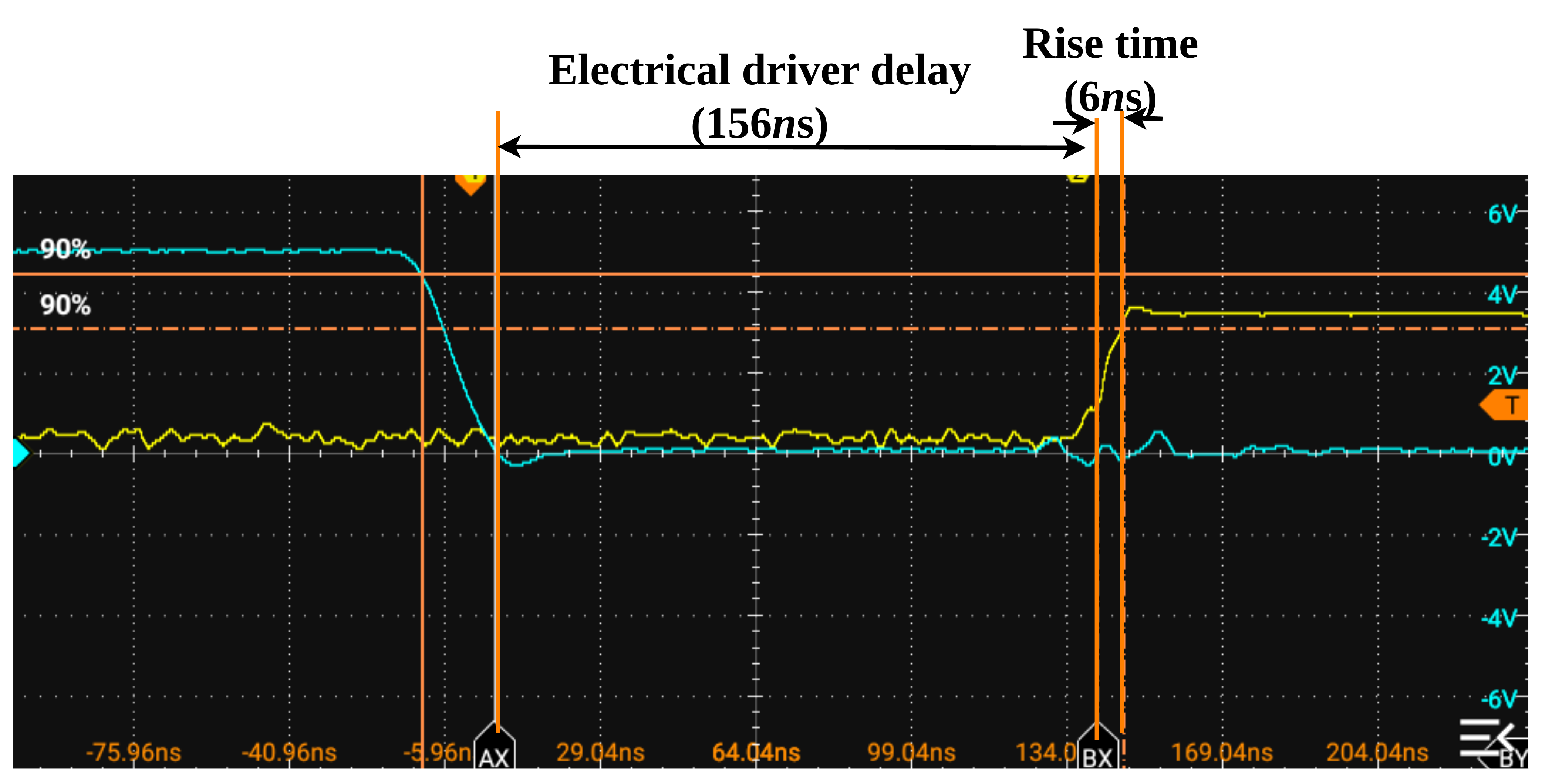}
\caption{Electrical fall to optical rise}
\end{subfigure}
\begin{subfigure}{0.45\linewidth}
\centering
\includegraphics[trim=1cm 0 1.5cm 0, clip,width=\linewidth]{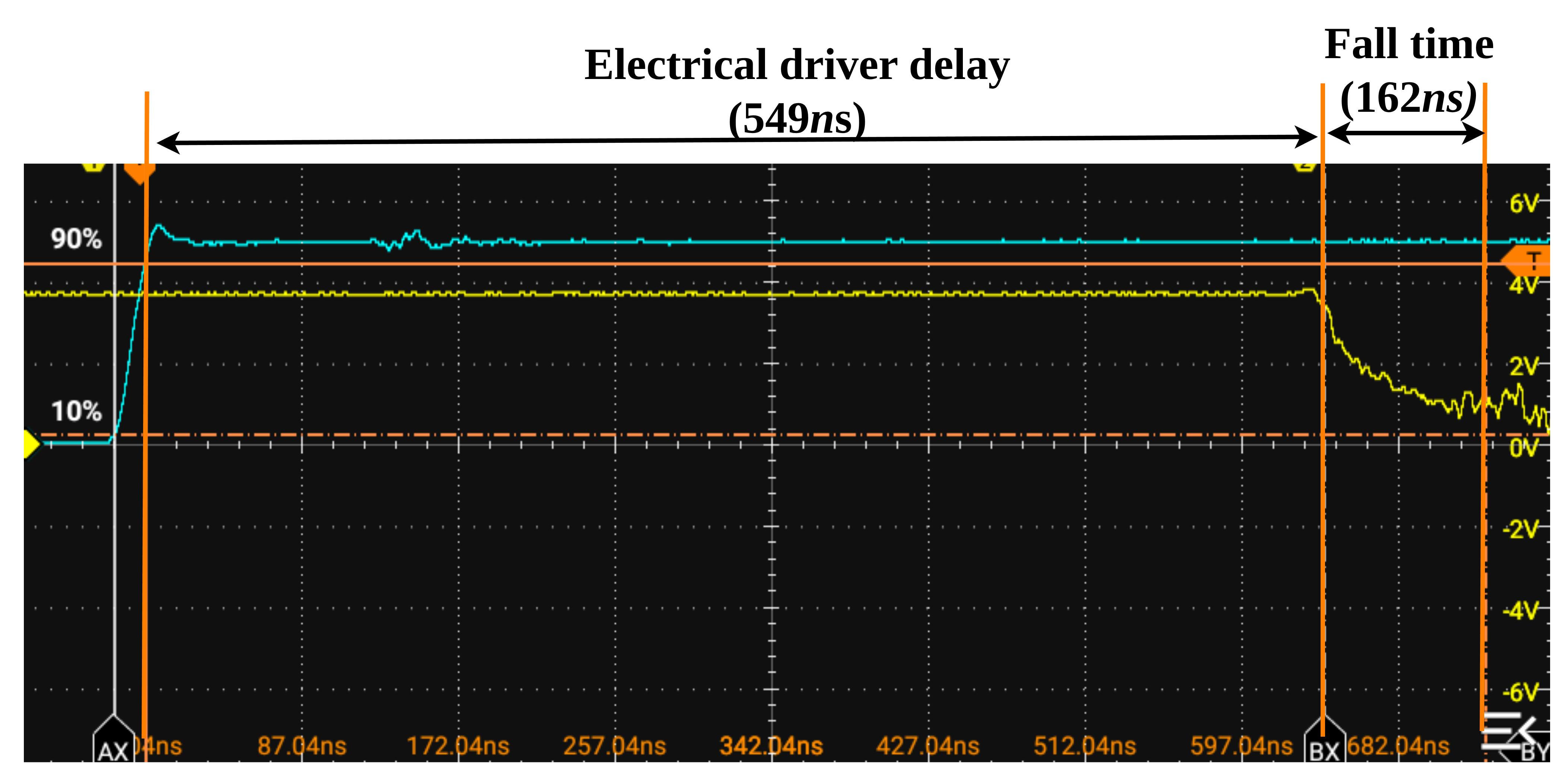}
\caption{Electrical rise to optical fall}
\end{subfigure}\hfill
\begin{subfigure}{0.45\linewidth}
\centering
\includegraphics[trim=0 0 0cm 0, clip,width=\linewidth]{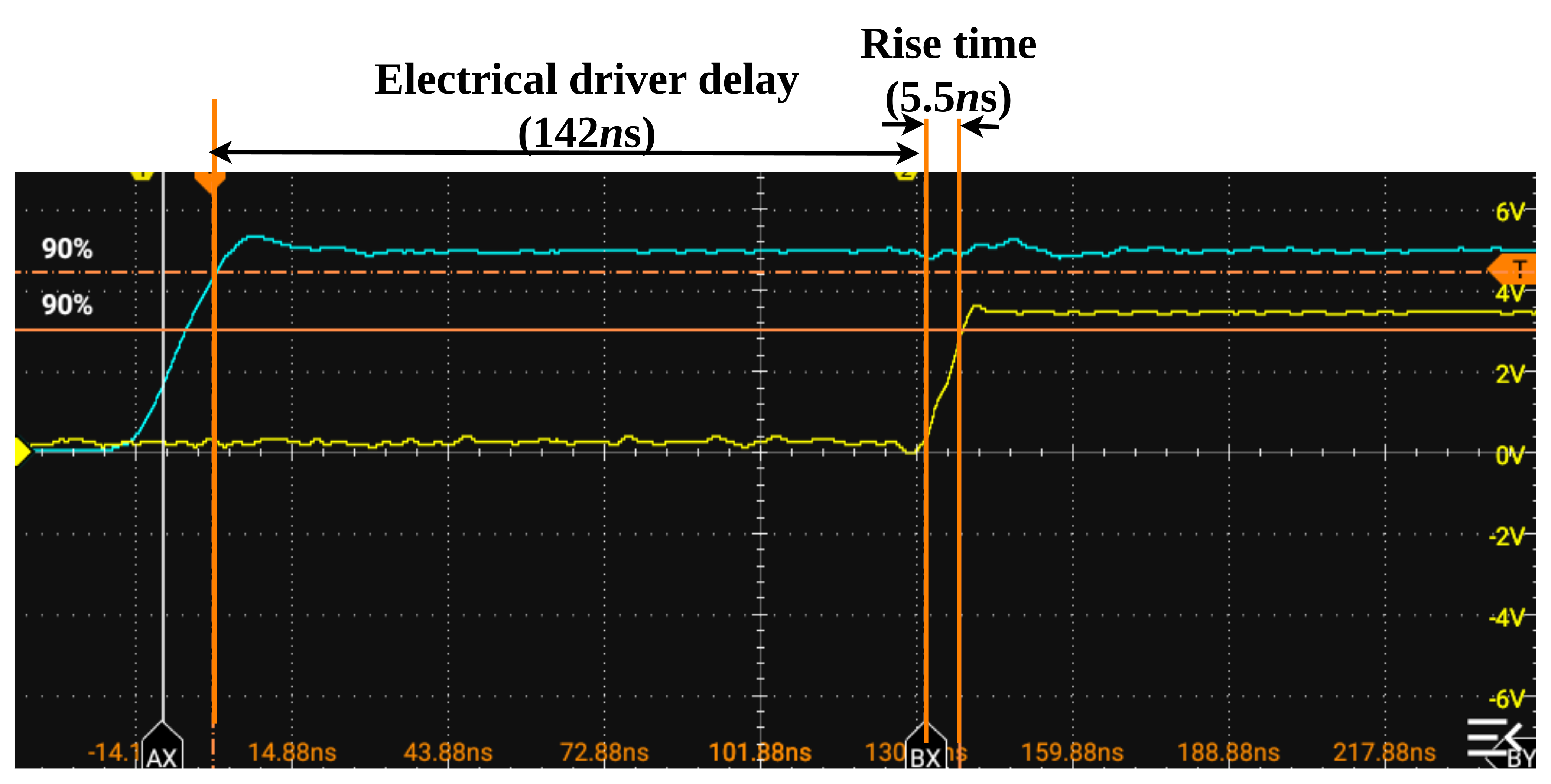}
\caption{Electrical rise to optical rise}
\end{subfigure}
\caption{Measured transition delays for a $2\times2$ optical switching element: While the intrinsic optical switching latency is approximately $150$\,ns, the electrical-to-optical response time reaches up to $600$\,ns, making the driver and actuation path a non-negligible contributor to the overall reconfiguration delay.}
\label{fig:switch-delay}

\end{figure*}

Our electro-optic switches operate at voltages substantially higher than the FPGA I/O levels. Each FPGA control signal first passes through a dedicated level translator before reaching the switch driver. The level translators contribute an additional $20$--$30$\,ns of latency, while the complete electrical signaling stage, including the driver circuitry, accounts for approximately $41\%$ of the measured end-to-end reconfiguration delay in our prototype. 

Figure~\ref{fig:switch-delay} shows the electrical-to-optical delay, defined as the time between issuing the electrical control signal and the onset of the optical transition, across the four possible switching transitions. We observe that the electrical-to-optical delay for optical rise transitions ranges from approximately $140$ to $160$\,ns, whereas the corresponding delay for optical fall transitions ranges from approximately $540$ to $600$\,ns. Figure~\ref{fig:rise-fall-variance} summarizes the distribution of these delays across all four transitions, showing that the electrical-to-optical delay exhibits non-negligible variation that should be accounted for when modeling nanosecond-scale photonic switching.

The driver circuitry also bounds the maximum reconfiguration frequency of the optical fabric. Even if the optical switching element itself operates in only a few nanoseconds, successive reconfigurations cannot occur faster than the driver can reliably actuate the switch. This observation extends beyond electro-optic devices: regardless of the underlying switching technology, the circuitry responsible for controlling the switching element forms an integral stage of the end-to-end reconfiguration pipeline.

\subsection{Optical Switching}

The final stage of reconfiguration is the optical transition itself. The latency of this stage is determined by the underlying switching mechanism. Free-space technologies such as 3-D MEMS~\cite{mems} and LCOS~\cite{LCOS} typically exhibit switching times ranging from microseconds to milliseconds, whereas integrated electro-optic devices achieve transitions in only a few nanoseconds~\cite{Phswi_datacntr,4x4}. Our prototype employs an electro-optic switch, enabling nanosecond-scale optical transitions.

Figure~\ref{fig:switch-delay} shows oscilloscope traces for the four possible state transitions of the $2\times2$ switch. Across all transitions, the optical rise and fall times remain between approximately $5$\,ns and $151$\,ns. The slight variation across transition types arises from the switching element's physical characteristics, but remains within the nanosecond regime. These measurements confirm that the optical transition itself contributes only a small fraction of the overall reconfiguration delay in our prototype. The dominant latency is outside the optical signal itself!

\subsection{Receiver Stack: This is the Not the End!}\label{subsec:receiver_latency}

The end-to-end reconfiguration pipeline does not end when the optical circuit is established. Once the optical transition completes, the receiver must reacquire transceiver lock, after which the NIC firmware, device driver, and host kernel react to the resulting link-state changes before communication can safely resume. These receiver-side effects dominate the overall reconfiguration delay and have been the subject of significant prior work~\cite{forencich2020system,xue2022nanosecond, whatocs}.

To understand this overhead, we measured the receiver stack latency using NVIDIA BlueField-3 NICs. Our testbed consistently exhibited approximately $1$\,s before the link became operational following a circuit transition. In contrast, kernel tracing showed that the Linux networking stack spends only $120$--$150\,\mu$s processing the corresponding link-state events. Since the \texttt{mlx5\_core} driver is open source while the NIC firmware remains proprietary, we cannot fully attribute the remaining latency. We  observed similar behavior on ConnectX-6 NICs. Prior work approaches receiver locking differently. NICs used in RotorNet’s \cite{10.1145/3651890.3672273} prototype lacked built-in low-latency loss-of-signal detection, allowing the switch reconfiguration events to be communicated to endpoints via an out-of-band notification channel. At ~20 µs slot times, this workaround was sufficient as locking overhead could be absorbed. Sirius \cite{sirius} eliminates the problem entirely through phase caching and storing the per-sender phase offset so that re-locking is nearly instantaneous, but this is only feasible because its fixed cyclic schedule makes the next sender’s identity known in advance.
Receiver recovery is intentionally outside the scope of the latency breakdown presented in Figure~\ref{fig:reconf_delay}. Our objective is to expose the often-overlooked control path that precedes receiver recovery. Nevertheless, once the control path itself is reduced to the microsecond scale, the receiver stack becomes the dominant component of application-visible reconfiguration delay and warrants renewed systems attention.

\section{Future Research Directions}\label{sec:future}

The objective of this paper is not merely to decompose reconfiguration delay, but to expose the systems challenges that emerge when photonic switching is viewed end-to-end. Our measurements suggest that reducing application-visible reconfiguration delay requires coordinated advances across the hardware--software stack. We highlight several research directions that naturally arise from this perspective.

\myitem{Hardware-efficient routing:}
Routing computation lies directly on the critical reconfiguration path, yet comparatively less attention has been paid to its absolute execution time on hardware than to its algorithmic complexity. Our measurements motivate routing algorithms that jointly optimize hardware execution latency, memory footprint, and resource utilization, alongside asymptotic complexity.

An alternative direction is to shrink the configuration space until precomputation becomes feasible again. Rather than supporting every possible circuit configuration, photonic interconnects could be designed around a carefully chosen subset of configurations that captures communication patterns observed in practice. Such restricted configuration spaces could dramatically simplify controller design and even enable lookup-table-based routing. Characterizing the resulting tradeoff between routing flexibility, controller complexity, and performance impact remains an interesting systems problem.

Oblivious switching~\cite{10.1145/3098822.3098838,10.1145/3651890.3672273,addanki2025vermilionl,10.1145/3579312,10.1145/3651890.3672248} represents one point in this design space, replacing online routing with a small set of precomputed permutations at the cost of multi-hop forwarding and synchronization. Understanding when demand-aware and oblivious switching provide the best end-to-end performance remains an open question.

\myitem{Multi-tenancy and interconnect sharing:}
Most prior work on circuit scheduling assumes either blocking interconnects, where every reconfiguration disrupts the entire fabric~\cite{10.1145/2486001.2486007,10.1145/1851182.1851223,10.1145/2716281.2836126,10.1145/3772356.3772395,Harvest,10.1145/2896377.2901479,9509349,10.1145/3748273.3749203,Scheduling_parallel_OCS}, or strictly non-blocking interconnects~\cite{10.1145/2999572.2999592}, where reconfigurations are independent across ports. Practical photonic interconnects, however, are often implemented as Rearrangeably Non-Blocking (RNB) networks, where reconfiguring one circuit may require modifying switches simultaneously used by other active connections. Consequently, a reconfiguration requested by one tenant can temporarily affect unrelated traffic sharing the same interconnect.

These interactions become increasingly important once reconfiguration is viewed as an end-to-end operation rather than a nanosecond-scale switching event. Circuit schedulers must reason not only about communication dependencies, but also about shared reconfiguration overheads arising from controller computation, electrical signaling, and receiver recovery. Developing scheduling algorithms that jointly optimize workload completion time while accounting for these interactions remains an important open systems problem. 

\myitem{Control interfaces for GPUs:}
Host-to-controller communication over PCIe accounts for a significant fraction of the end-to-end reconfiguration delay despite transferring only a few bytes. This raises a simple question: is a general-purpose memory transaction the right abstraction for circuit control? Circuit reconfiguration requires only reliable command delivery, ordering, and completion semantics. Identifying the minimal hardware--software protocol that provides these guarantees while minimizing latency presents an interesting systems research direction.

\myitem{Receiver stack and NIC support:} The receiver stack and transceiver recovery remain the dominant contributors to application-visible reconfiguration delay. Significant prior work has reduced transceiver locking to nanosecond-scale using burst-mode transceivers~\cite{forencich2020system} and multicast-based control protocols reporting $3.1$\,ns clock and data recovery~\cite{xue2022nanosecond}. While this is only one milestone in the reconfiguration process, the application can resume communication only after the receiver stack, NIC, and operating system collectively recognize that the newly established circuit is ready~\cite{whatocs}. This exposes a broader systems challenge: what does it mean for a circuit reconfiguration to be \emph{complete}? The controller, optical switch, transceiver, NIC, operating system, and application each observe different completion events. Defining the appropriate completion semantics across these layers, and determining how these events should be propagated to the application, presents an interesting systems research direction.

\myitem{Generalization of the measurements:} The absolute latencies we report are specific to our prototype, but which stages lie on the critical path depends on how reconfiguration is controlled. \emph{PCIe write:} In GPU-controlled reconfiguration, for example, at the boundary of a collective communication step, the circuit request must travel from the GPU to the controller. A GPU-issued PCIe write to a discrete controller (\S\ref{pcie}) is the most direct such path. Passing the request through the CPU only increases this cost, since the GPU must first notify the CPU before the write is issued. Conversely, dedicated control interfaces for GPUs, as discussed above, may provide a more direct, lower-latency channel between GPUs and the controller. In switch-centric reconfiguration, where the switch controller follows a fixed schedule~\cite{10.1145/3098822.3098838,sirius}, no per-request message is needed, and this stage disappears. \emph{Routing:} Routing is needed only when the configuration must be computed at runtime; fixed configurations can be precomputed, as discussed in \S\ref{routing} and under \emph{Hardware-efficient routing} in \S\ref{sec:future}. \emph{Signaling:} Every photonic switch must be actuated electrically, so signaling is present in all designs. However, the latency depends on the switch's actuation mechanism, whether thermo-optic, electro-optic, magneto-optic, or MEMS-based (\S\ref{subsec:elecdriver}).

We believe end-to-end photonic switching has emerged as a rich systems problem. As switching technologies and circuit scheduling algorithms continue to mature, the next performance gains will come from optimizing the end-to-end reconfiguration pipeline.

\section*{Acknowledgements}
We thank the anonymous reviewers and our shepherd for their valuable feedback. This work was supported by the National Science Foundation (NSF) under Grant IIS-2551372.

\label{bodyLastPage}

\bibliographystyle{plain}
\bibliography{reference}

\label{LastPage}
\end{document}

%% file: macros.tex
\usepackage{enumitem}

\makeatletter
 \renewcommand\subsubsection{\@startsection {subsubsection}{3}{\z@}%
     {-2.5ex \@plus -1ex \@minus -.2ex}%
     {1.3ex \@plus.2ex}%
    {\ACM@NRadjust\@subsubsecfont}}
\makeatother

\definecolor{darkgreen}{rgb}{0.01, 0.75, 0.24}
\definecolor{darkcyan}{rgb}{0.0, 0.55, 0.55}
\definecolor{mycolor}{rgb}{0.01, 0.75, 0.24}
\definecolor{takeawaycolor}{rgb}{0.75, 0, 0}
\definecolor{mygray}{gray}{0.9}
\definecolor{myred}{rgb}{0.75, 0, 0}
\definecolor{alg1}{rgb}{0.898, 1, 0.914}
\definecolor{alg2}{rgb}{0.925, 0.988, 1}
\definecolor{alg3}{rgb}{1, 1, 0.906}
\definecolor{alg4}{rgb}{1, 0.922, 0.922}
\mathchardef\hyphen="2D

\newtheoremstyle{sltheorem}
{}                %
{}                %
{\slshape}        %
{}                %
{\bfseries}       %
{.}               %
{ }               %
{}                %
\theoremstyle{sltheorem}

\newmdenv[
  backgroundcolor=gray!20,
  linecolor=gray,
  linewidth=1pt,
  roundcorner=5pt,
  skipabove=6pt,
  skipbelow=6pt,
  innertopmargin=1pt,
  innerbottommargin=4pt,
  innerleftmargin=6pt,   %
  innerrightmargin=6pt   %
]{graybox}

\newcommand{\myitem}[1]{\vspace{1mm}\noindent\textbf{#1}}

\newcommand{\remove}[1]{}

\newcommand{\first}{\emph{(i)}\xspace}
\newcommand{\second}{\emph{(ii)}\xspace}

\usepackage{xcolor}
\definecolor{light-gray}{gray}{0.9}
\usepackage{tcolorbox}

\newcommand{\cref}[1]{Chapter~\ref{#1}}

\newcommand*\circled[1]{\tikz[baseline=(char.base)]{
            \node[scale=0.9,shape=circle,draw,inner sep=3.5pt] (char) {};
            \node[scale=1.0] (char) {\textcolor{black}{\footnotesize{\textbf{#1}}}}}}

\newcommand{\benes}{Bene\v{s}\xspace}